\documentclass{aa}  

\usepackage{graphicx}
\usepackage{txfonts}
\usepackage[usenames]{xcolor}
\usepackage[normalem]{ulem}

\def\kms{{\rm km/s}}
\def\feh{{\rm [Fe/H]}}

\begin{document}


\title{Binary-single interactions with different mass ratios}
 \titlerunning{binary-single interactions}

\subtitle{Implications for gravitational waves from globular clusters}

   \author{Bruno Rando Forastier
          \inst{1,2}\thanks{bruno.randof@gmail.com},
          Daniel Mar\'{i}n Pina,
          \inst{1}
          Mark Gieles
          \inst{1,3},
          Simon Portegies Zwart
          \inst{2},
          Fabio Antonini
          \inst{4}
          }

   \institute{Institut de Ciències del Cosmos (ICCUB), Universitat de Barcelona (IEEC-UB),
              Martí Franquès 1, E08028 Barcelona
         \and
             Leiden Observatory, University of Leiden, Niels Bohrweg 2, 2333 CA Leiden
              \and
              ICREA, Pg. Llu\'{i}s Companys 23, E08010 Barcelona, Spain
              \and 
              Gravity Exploration Institute, School of Physics and Astronomy, Cardiff University, CF24 3AA, UK
             }
\authorrunning{Rando et al.}
   \date{Received XXXXX; accepted YYYYY}

 
  \abstract
{
Dynamical interactions in star clusters are an efficient mechanism to produce the coalescing binary black holes (BBHs) that have been detected with gravitational waves (GWs). 
}
{ 
We want to understand how BBH coalescence can occur during -- or after --  binary-single interactions with different mass ratios.
}
{ 
We perform gravitational scattering experiments of binary-single  interactions using different mass ratios of the binary components ($q_2\equiv m_2/m_1\le1$) and the incoming single  ($q_3\equiv m_3/m_1$). We extract cross sections and rates for (i) GW capture during resonant interactions; (ii) GW inspiral in between resonant interactions and apply the results to different globular cluster conditions. 
}
{ 
We find that GW capture during resonant interactions is most efficient if $q_2\simeq q_3$ and that the mass-ratio distribution of BBH coalescence due to inspirals is
$\propto m_1^{-1}q^{2.9+\alpha}$, where $\alpha$ is the exponent of the BH mass function. The total rate of GW captures and inspirals depends mostly on $m_1$ and is relatively insensitive to $q_2$ and $q_3$.
We show that eccentricity increase by non-resonant encounters approximately doubles the rate of BBH inspiral in between resonant encounters. {For a given GC mass and radius, the BBH merger rate in metal-rich GCs is approximately double that of metal-poor GCs, because of their (on average) lower BH masses ($m_1$) and steeper BH mass function, yielding  binaries with lower $q$.
}
}
{ 
{Our results enable the translating from the mass-ratio distribution of dynamically formed BBH mergers to the underlying BH mass function.
The additional mechanism that leads to a doubling of the inspirals provides an explanation for the reported high fraction of in-cluster inspirals in $N$-body models of clusters.    }
}

   \maketitle{}
%

\section{Introduction}

The first binary black hole (BBH) coalescence was detected in 2015 by LIGO (the Laser Interferometer Gravitational Wave Observatory) \citep{2016LIGO}. In the first three observing runs of the LIGO-Virgo-Kagra (LVK) Collaboration, a total of 90 compact binary coalescences (CBCs) were reported,  involving  black holes (BHs) and neutron stars (NSs) \citep{2021LVK}. At the time of writing, around 80 more were found in O4a. Gravitational wave (GW) astronomy allows us to observe the Universe from a completely new perspective, unveiling phenomena that remain invisible to the traditional electromagnetic observations. The dominance of BHs involved CBCs has raised the question of how these binaries form and eventually inspiral. The literature on this topic is rich and multiple formation scenarios have been proposed for CBC. We refer  to the review by \cite{mandel_2022} and references therein for an overview. Important for this work is that -- especially for massive BBHs -- dynamical assembly has been suggested to be an efficient mechanism for creating GW sources \citep[e.g.,][]{2000ApJ...528L..17P,2015PhRvL.115e1101R, 2018MNRAS.481.5123B,dicarlo19}.

In this paper, we are concerned with the dynamical interactions of BHs that lead to mergers within the cores of globular clusters (GCs). In GCs, there is a large number of stellar remnants such as BHs and NSs, which are formed during the early stages of the cluster, when the  massive stars end their lives. Those compact remnants sink towards the centre of the cluster due to dynamical friction. It is in this environment where interactions are frequent, and the formation of BBHs is  driven by the energy need for the relaxation of the GC \citep{henon_1975,breen_heggie_2013}. Tight binaries harden as the result of interactions \citep{heggie_1975}, and give away part of their energy to their surroundings in the form of kinetic energy. When enough energy is given away, the BBHs can merge through the emission of GWs.  It is believed that these binary systems could be responsible for a significant fraction of the massive mergers identified by the LVK collaboration to date \citep[e.g.,][]{2016PhRvD..93h4029R,2023MNRAS.522..466A}. The interactions that drive the hardening process of the BBH can be binary-single encounters, or higher order configurations such as binary-binary encounters \citep{zevin_2019}. 

The topic of this study is  the eventual coalescences of BBHs in GCs through binary-single encounters. Particularly, we conduct scattering experiments as in \citet[][]{hut_bahcall_1983}, but with unequal component masses. Because BHs of  different masses coexist and interact in GCs, the inclusion of different masses in our simulations  allows us to better understand the phenomena surrounding these BH interactions. For example, including arbitrary masses can shed light into the mass-ratio distribution from mergers detected by the LVK collaboration. {To better represent binary-single encounters in GCs, we weigh our scattering events according to the mass distributions of the three bodies, which are obtained from a model for BBH coalescences in GCs \citep[as in][]{antonini_gieles_2020}.} Although useful, the unequal-mass regime further complicates the already complex equal-mass three-body problem. It is for this reason that we use Newtonian dynamics: this allows us to determine the dynamics of the system not by the absolute masses, but rather by the mass ratios of the BHs, reducing the dimension of the parameter space by one and simplifying the analysis. The elephant in the room is that we aim at understanding the physics surrounding BH mergers with Newtonian dynamics, which at first glance appears to be contradictory. The work of \cite{samsing_2014,samsing_2018,samsing_2020}, \cite{antonini_gieles_2020}  shows that the rate of inspiral in post-Newtonian scattering experiments can be well described by simple Newtonian theory using a minimum distance/eccentricity criterion for inspiral to occurs. We can, therefore, determine a posteriori which of the Newtonian interactions would lead to mergers.

In parallel to the GW implications of binary-single encounters,  we also use our unequal-mass simulations to revisit some generalities of the three-body problem. Extensive work has been done in the equal-mass case \citep{hut_bahcall_1983,heggie_hut_1993}, as well as in the unequal-mass case \citep{hills_fullerton_1980,spitzer_1987,sigurdsson_1993,heggie_mcmillan_hut_1996,quinlan_1996}. The main work has been put into the different outcomes that can result from a binary-single encounter, such as exchanges of one of the components, close approaches, and eccentricity or energy changes in the initial binary. With the growing interest in GWs, later literature on binary-single interactions also includes the possibility for CBCs \citep{gultekin_2006,samsing_2014,ginat_2022}. We build upon their work by exploring some yet unknown intricacies of resonant interactions. As we will explain later on, during a resonant binary-single interaction, a bound but unstable three-body system is formed, where the three components approach each other repeatedly. After such an encounter, the original binary usually increases its binding energy, and its eccentricity can be completely different than its initial value. We study how this energy change, the final eccentricity distribution, and the number of repeated approaches depend on the masses of the triplet.

This paper is organised as follows: in Section \ref{methods} the terminology and settings of the binary-single scattering problem are explained. We also describe the initial conditions required to fully determine a scattering experiment, followed by a description of the software package used to run the simulations. In Section \ref{results 1} we present  general results of the binary-single scattering experiments. In Section \ref{model} we apply the results of the scattering experiments to different GC conditions. In Section \ref{results 2} we focus our analysis to the specific case of BBH coalescences happening in GCs. The limitations of our analysis are discussed in Section \ref{sec discussion}, and our findings are summarised in Section \ref{conclusions}.

\section{Methods: scattering experiments}
\label{methods}

\subsection{The binary-single scattering problem}
\label{overview}
Binary-single scattering refers to the specific scenario where a binary interacts gravitationally with a single body. This work seeks to understand the dynamics, outcomes, and effects of these interactions, with a particular focus on encounters involving BHs of different masses. The simulations are done in the Newtonian point-mass limit. It is for this reason that in this Section and in Section \ref{results 1}, the point masses will be referred to as bodies. From now on, the components of the initial binary are labeled as 1 and 2, with 1 being the heaviest, and the single body is labeled as 3.\\ 

The terminology used in the gravitational scattering problem has some analogies with particle physics experiments. When the final three-body system has no bound states between bodies, we say that the system is ionized. Furthermore, when the final binary is composed of two different bodies than the initial binary, we refer to the outcome as an exchange, while a preservation refers to a final binary with the same components as the initial binary. More analogies with particle physics are found in Section \ref{cross sections}, where we discuss the use of cross sections. 

The gravitational three-body problem is in almost all cases chaotic, and some of the scatterings can allow for temporarily bound three-body systems: we call these interactions resonant. A way to clearly define whether an interaction is resonant is through the number of minima in $s^2$ as a function of time, defined as
\begin{equation}
    s^2 \equiv r_{12}^2 + r_{23}^2 + r_{13}^2 , 
\end{equation}
where $r_{ij}$ is the distance between  bodies $i$ and $j$. For very early and very late times, $s^2$ is arbitrarily large, because at least one of the three bodies is at infinite distance from the binary. During the closest interaction, $s^2$ can have either one or multiple minima: this number of minima is called the number of intermediate states $N_{\rm IMS}$. If the interaction is such that $N_{\rm IMS} > 1$, we label it as resonant. Otherwise, if $s^2$ has a single minimum ($N_{\rm IMS} = 1$) the interaction is non resonant, or direct \citep{heggie_mcmillan_hut_1996}. We consider two consecutive minima to be distinct  if the value of $s^2$ at the maximum is at least twice that of the minima.

Among resonant interactions, we  distinguish democratic resonances from hierarchical resonances \citep{heggie_hut_1993}. Democratic resonances are those that do not favour the interactions of a single pair of bodies with respect to the other two pairs, while hierarchical resonances are characterized by having a pair of bodies strongly interacting, with the third body on a wider orbit around them. Even though some times it is easily seen which kind of resonance corresponds to a given interaction, the exact line that separates democratic and hierarchical is loose. For example, in nature some resonant interactions will live enough to oscillate from democratic to hierarchical (and vice versa), making the classification problematic. It is for this reason that we are not separating on this classification, but it is worth mentioning, to illustrate the richness and depth of the gravitational scattering problem.

Moreover, one may also consider the formation of stable three-body systems. If systems like the Sun-Earth-Moon exist, then why are we not considering them as possible outcomes of a scattering? The answer to this question is that scattering events are unable to form stable and bound three-body systems, because the set of initial conditions that lead to a stable three-body system is of measure zero with respect to the initial hyperparameter space. A rigorous proof can be found in \cite{chazy_1929}, but \cite{hut_bahcall_1983} gives an intuition for this unlikeliness. A three-body stable system that  formed from a binary-single scattering event should be stable for the infinite future. But, if we consider Newton's time reversibility, we expect the system to be stable for the infinite past, which contradicts the assumption of the system being created by a scattering event.

Our methods to identify GW captures and inspirals in our Newtonian simulations are defined in Section \ref{GW captures} and Section \ref{inspirals}, respectively. For extended bodies, an additional possible outcome is a direct collision \citep[for stellar collisions, see e.g.][]{2004MNRAS.352....1F}. The cross section (defined in Section \ref{cross sections}) for a collision between NSs is orders of magnitude lower  than that of the other types of mergers \citep{samsing_2014}, and because we only consider BHs,  we do not  consider collisions in this work.

\subsection{Description of the code and units}
\label{STARLAB}

\begin{table}
\caption{Parameters needed to specify the initial conditions of a scattering simulation, including their corresponding ranges and the distributions.}
\label{parameters}
\centering
\begin{tabular}{l l p{3.5cm} l}
\hline\hline
Param. & Values & Description & Distribution \\
\hline
$q_2$ & [0.1, 1] & Mass of body 2 & Uniform \\
$q_3$ & [0.1, 1] & Mass of body 3 & Uniform \\
$e$ & [0, 1) & Initial eccentricity & $e^2$ \\
$\tilde{v}$ & 0.1 & Relative velocity of body 3 & ... \\
$b$ & [0, $b_{\rm max}$) & Impact parameter & $b^2$ \\
$\phi$ & [0, 2$\pi$) & Phase of binary & $\phi$ \\
$\theta$ & [0, $\pi$/2] & Angle of incidence & $\cos\theta$ \\
$\psi$ & [0, 2$\pi$) & Orientation of body 3 & $\psi$ \\
$f$ & [0, 2$\pi$) & Projected true anomaly & $E'-e\sin E'$ \\
\hline
\end{tabular}
\end{table}

To simulate the binary-single scattering events, we use the \textsc{starlab} software package. \textsc{starlab} is a collection of  tools designed to simulate and analyze the dynamical evolution of star clusters, which is described in \cite{mcmillan_hut_1996_starlab}. Within \textsc{starlab}, the module \textsc{scatter3} simulates  binary-single scattering events using Newtonian physics. It uses a fourth-order variable time step Hermite integrator (described in \citealt{makino_1992}). 
To keep integration errors under control,  \textsc{scatter3} uses a technique developed in \cite{1995leapfrog} that ensures the symmetry in time, which guarantees that the energy and momentum are kept constant in periodic orbits.

Part of the complexity of the three-body problem is the amount of parameters that need to be specified in order to fully determine the initial conditions. In the point-mass limit, each body requires information of its position,  velocity and mass, a total of seven parameters. Therefore, three bodies  require 21 parameters. Thankfully, the symmetries of the system allow for a considerable reduction in the number of parameters. In addition, the scale invariance in the physical dimensions of length, time and mass allow for a further reduction of parameters.

We adopt units in which the semi-major axis (SMA) $a$ of the initial binary, the gravitational constant $ G$, and the mass of the initial binary $m_{12}$ are  ${ G}=a=m_{12}=1$. The initial conditions required to simulate a binary-single scattering can be reduced to  nine parameters, given in Table~\ref{parameters} \citep{hut_bahcall_1983}. The impact parameter is given in units of $a$, and the velocity $v$ is the relative velocity between the binary and the field BH at infinity. We define $\tilde{v}$ as the velocity in units of the critical velocity $v_{\rm c}$, defined as the relative velocity that makes the energy of the three-body system to be zero:
\begin{equation}
\label{v_c}
     v_{\rm c}^2 = { G}\frac{m_1m_2m_{123}}{m_3m_{12}}\frac{1}{a} , 
\end{equation}
with $m_{123}\equiv m_{12}+ m_3$. Note that if $\tilde{v}<1$, the energy of the three-body system is negative, which implies that ionization is physically impossible. If $\tilde{v}>1$, the energy of the system is positive. Therefore, bound three-body systems are not permitted, and resonant interactions do not happen.

For this work, we define the mass ratios $q_2\equiv m_2/m_1$ and $q_3\equiv m_3/m_1$. We explore different values of $q_2$ and $q_3$ in the range $[0.1,1]$, spaced in steps of 0.05. For each combination of $q_2$ and $q_3$, 31250 encounters were simulated (for a total of $\mathcal{O}(10^7)$ encounters). This order of magnitude was selected to make the statistical error of mergers in each bin sufficiently low. Following Table~\ref{parameters}, the encounters were simulated with eccentricities sampled from a thermal distribution $p_{e}(e)=2e$ \citep{jeans_1919}, randomly sampled angles, a velocity $v=0.1$, and impact parameters equispaced in $b^2$ in the range $b\in[0,b_{\rm max}]$. The maximum impact parameter $b_{\rm max}$ is chosen heuristically by imposing the condition that no resonant encounters happen for $b\gtrsim0.6b_{\rm max}$. Less than 0.1\% of the encounters were discarded, either because they were unresolved or because $N_{\rm IMS} > 500$\footnote{These are a minority of encounters and they require a vast amount of integration steps, and are prone to accumulating errors. We therefore discard them.}.

\subsection{Cross sections}
\label{cross sections}

In this Section we present the physical quantities that are of interest for the scattering problem. First, we define $\mathcal{E}$ as the absolute value of the binding energy of the binary
\begin{equation}
    \mathcal{E} = \frac{{G}m_1m_2}{2a} .
\label{bin}
\end{equation} 

This binding energy can increase or decrease through interactions with a third body. The relative change in $\mathcal{E}$ is $\Delta$, defined as
\begin{equation}
    \Delta \equiv \frac{\mathcal{E}(t\xrightarrow{}\infty) - \mathcal{E}(t\xrightarrow{}-\infty)}{\mathcal{E}(t\xrightarrow{}-\infty)} .
\end{equation}
Note that $\Delta$ goes from $-1$, corresponding to binary ionization, to $\infty$, implying the collapse of the binary. We define the cross section for an event X to happen as
\begin{equation}
\label{sigma}
    \Sigma_{\rm X} = {\pi}b_{\rm max}^2P_{\rm X},
\end{equation}
where $P_{\rm X}$ is the fraction of interactions with an impact parameter $b<b_{\rm max}$ that undergo an outcome X (e.g., resonant, exchange, etc.). When $b_{\rm max}\rightarrow\infty$, nearly all interactions are distant flybys that do not perturb the binary, making $P_{\rm X}\rightarrow 0$  such that $\Sigma_{\rm X}$ converges\footnote{As long as the outcome X is not possible above a finite impact parameter, $\Sigma_{\rm X}$ converges. This does not happen with, for example, flybys, which are allowed for any arbitrary impact parameter.}. In practice, $b_{\rm max}$ is an infrared cutoff chosen to be sufficiently large so that $\Sigma_{\rm X}$ converges. We describe our choice of $b_{\rm max}$ in Section \ref{STARLAB}. Following the same argument, $\langle\Delta\rangle$ is a quantity that depends on the arbitrary choice of $b_{\rm max}$, making it an ill-defined quantity. From now on, $\Delta_{\rm r}$ refers to the relative change in energy in resonant interactions, whose mean value is a well-defined quantity. Furthermore, given a cross section $\Sigma_{\rm X}$, one can compute the rate $R_{\rm X}$ at which encounters of type X happen between the binary and its surrounding bodies: 
\begin{equation}
\label{eq rates}
    R_{\rm X} = n\langle v\Sigma_{\rm X}\rangle .
\end{equation}
where $n$ is the density of the surrounding bodies.

When computing cross sections, there are two sources of errors \citep{hut_bahcall_1983}. The first is the statistical error present in all probabilistic processes, given by
\begin{equation}
   \delta_{\rm stat}\Sigma_{\rm X} = \frac{\Sigma_{\rm X}}{\sqrt{N_{\rm X}}},
\end{equation}
where $N_{\rm X}$ is the number of interactions of type X. The second source of error comes from the computation: to keep in check this source of numerical inaccuracies, the integration steps must be kept small enough. As it is stated in Section \ref{STARLAB}, the error in the total energy of the system is kept under control, hence making the systematic errors that come from numerical inaccuracies negligible, much less than 1 \% \citep{mcmillan_hut_1996_starlab}. It is difficult to have a measure of the computational error, so we conservatively assume it is equal to the statistical error. In addition, our definition of a minimum in $s^2$ (Section \ref{overview}) is a source of error when classifying events as resonant or non resonant. This source of error is again difficult to estimate analytically, and as a first approximation it is considered to be within the computational error.

\section{Results: generalities of unequal-mass encounters}
\label{results 1}

In this Section we present results obtained from our binary-single scattering events with unequal-masses. These results are not restricted to BHs, but are general for all binary-single scatterings. The focus is on understanding how the behavior of resonant interactions varies with the masses of the different bodies.

\subsection{$\langle\Delta_{\rm r}\rangle$ for arbitrary masses}

   \begin{figure}
   \centering
   \includegraphics[width=\linewidth]{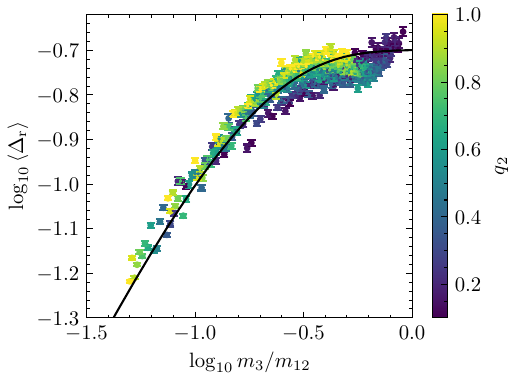}
      \caption{ $\langle\Delta_{\rm r}\rangle$, in units of $\Delta_0$, as a function of $m_3/m_{12}$. Each data point represents a different combination of $q_2$ and $q_3$. The solid line is the fitted function.
              }
         \label{delta m3}
   \end{figure}

It is generally assumed that during a binary-single resonant encounter, the binary increases its binding energy by around a 20\% \citep{spitzer_1987,heggie_hut_1993,quinlan_1996,portegies_2010}. Nonetheless, the validity of this result is restricted to the equal-mass case. In Fig.~\ref{delta m3} we show how $\langle\Delta_{\rm r}\rangle$ behaves as a function of $m_3/m_{12}$, where we have realised the fit
\begin{equation}
\label{eq delta m3}
   \langle \Delta_{\rm r}\rangle \simeq  \Delta_0\left[1- {\rm exp}\left(-A\frac{m_3}{m_{12}}\right)\right] ,
\end{equation}
with $\Delta_0 \equiv\langle\Delta_{\rm r}\rangle\left(m_1=m_2=m_3\right)=0.2$ and $A=7.0$. This fit is restricted to $m_3\leq m_{12}$. From now on, equation (\ref{eq delta m3}) is used for all calculations that require $\Delta_{\rm r}$. We choose as x-axis the ratio $m_3/m_{12}$ rather than $q_3$, because for the latter the data points disperse at low values of $q_3$.

Although our results agree qualitatively with what \cite{hills_fullerton_1980} found, our values of $\langle\Delta_{\rm r}\rangle$ are  a factor of $\sim2$ higher for $m_3/m_{12}\lesssim0.1$ and a factor of $\sim4$ lower for $m_3/m_{12}\sim 1$. We note that their encounters were head-on, and we can reproduce their results with our runs with $b=0$. These head-on events are unrealistic, and they are physically different than more realistic scatterings with $b\neq0$ due to the difference in angular momentum and in $\langle\Delta_{\rm r}\rangle$.

A dynamically-assembled BBH in a GC tends to contain the two most massive BHs in the cluster. The eventual binary-single encounters happen with the binary's stellar surroundings, where bodies generally have masses below $m_{12}$. In the case of interactions with stars, the difference in mass can even reach two orders of magnitude. In this scenario, equation (\ref{eq delta m3}) implies that $\langle\Delta_{\rm r}\rangle$ is generally lower than 0.2. For the BH mass functions assumed in Section~\ref{model} we find  typical values of $\Delta_{\rm r}\simeq0.15-0.17$, not very different from the equal-mass value of $0.2$.

\subsection{Number of intermediate states}
\label{subsection nims}

   \begin{figure*}
   \includegraphics[width=0.5\linewidth]{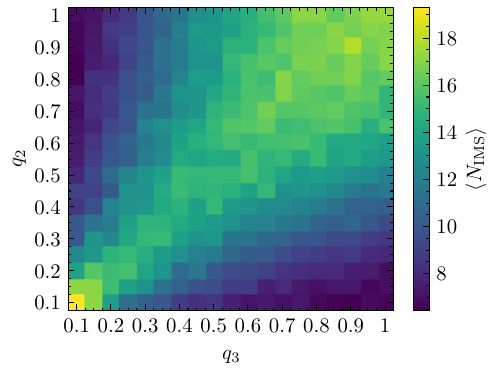}
    \includegraphics[width=0.5\linewidth]{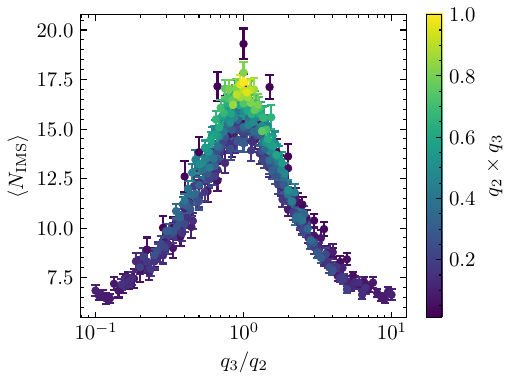}
              

  
      \caption{{\it Left:} $\langle N_{\rm IMS}\rangle$ for resonant encounters with different mass ratios.
             {\it Right:} $\langle N_{\rm IMS}\rangle $ for resonant encounters, as a function of $q_3/q_2$. The error bars correspond to $\sqrt{2}$ times the standard error of the mean.
              }
         \label{Nims}
   \end{figure*}

   \begin{figure*}
   \includegraphics[width=0.5\linewidth]{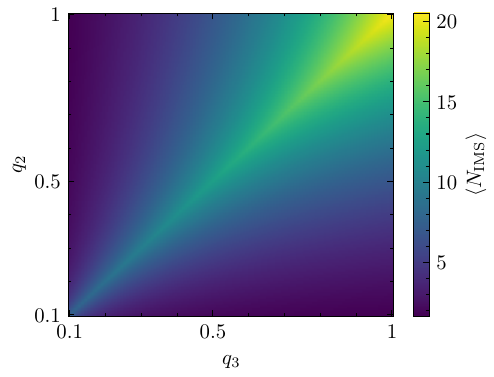}
    \includegraphics[width=0.5\linewidth]{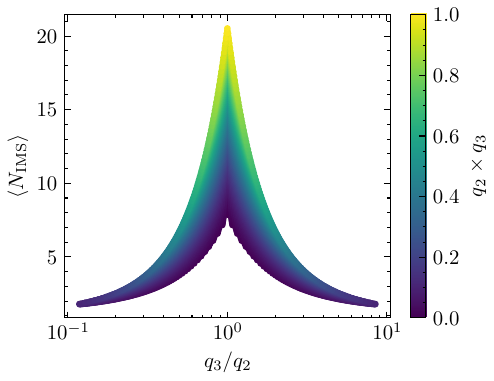}
              

  
      \caption{{\it Left:} Theoretical $\langle N_{\rm IMS}\rangle$ for resonant encounters with different mass ratios.
             {\it Right:} Theoretical $\langle N_{\rm IMS}\rangle $ for resonant encounters, as a function of $q_3/q_2$.
              }
         \label{Nims theo}
   \end{figure*}


The number of intermediate states in a resonant encounter, $N_{\rm IMS}$, is crucial, as the GW capture probability depends linearly on it (see Section \ref{results 2}). For equal masses, $\langle N_{\rm IMS} \rangle \simeq 20$ \citep[][]{samsing_2018}, but its behavior for unequal masses remains unknown.

In Fig.~\ref{Nims} we show how $\langle N_{\rm IMS}\rangle$ behaves in the unequal-mass case. This quantity is maximised in the region where $m_2\simeq m_3$, which also includes the equal-mass case. In this region, $\langle N_{\rm IMS}\rangle \simeq 20$, but outside of it $\langle N_{\rm IMS}\rangle$ drops considerably.

We can intuitively understand this as follows: if one of the bodies is  less massive than the other two, then that body is most likely to be ejected \citep{heggie_1975,sigurdsson_1993,heggie_mcmillan_hut_1996}. The probability of ejection per intermediate state is high and fewer IMSs are, therefore, needed to end an resonant interaction. Meanwhile, if $m_3\simeq m_2$, the probability of ejection per IMS is lower because there is no obvious candidate for ejection and more IMSs are required to end the interaction. The highest number is for equal masses, when the interaction is  democratic.

We  now present a more rigorous explanation for the behaviour of  $N_{\rm IMS}$. In this derivation, we follow the derivation found in \cite{samsing_2018} and \cite{samsing_2024}, but generalising it to arbitrary masses (some steps are explained more in detail in the referenced papers). Let us begin by decomposing the triplet during one IMS into a temporary binary of SMA $a_{\rm IMS}$ and (positive) binding energy $\mathcal{E}_{\rm IMS}$, and a single body that orbits the binary on a wider orbit. We assume that, during each IMS, the energy of the binary is randomly sampled from a distribution $p_{\mathcal{E}}(\mathcal{E}_{\rm IMS})\propto \mathcal{E}_{\rm IMS}^{-\gamma}$, where we specify the value of $\gamma$ later on. In the hard binary limit, if $\mathcal{E}_{\rm IMS}$ is higher than the energy of the initial binary $\mathcal{E}_0$, then the remaining energy necessarily goes to the single, and the interaction ends with an ejection. In the other hand, if $\mathcal{E}_{\rm IMS}< \mathcal{E}_0$, the binary absorbs part of the single's energy and it remains bound until the next IMS. Then, $N_{\rm IMS}$ can be approximated by \citep{samsing_2018}
\begin{equation}
\label{eq Nims basic}
    \langle N_{\rm IMS}\rangle \simeq \frac{P(\mathcal{E}_{\rm IMS}<\mathcal{E}_0) }{P(\mathcal{E}_{\rm IMS}>\mathcal{E}_0)}\simeq \left(\frac{a_{\rm max}}{a_0}\right)^{\gamma-1},
\end{equation}
where $a_{\rm max}$ is the maximum value that $a_{\rm IMS}$ can have, or equivalently, the SMA at which $\mathcal{E}_{\rm IMS}$ is minimised. Consider the case $q_3 < q_2$, and we assume that the temporary binary is always composed by the two heaviest bodies in the triplet (in this particular case, bodies 1 and 2). $\mathcal{E}_{\rm IMS}$ attains its minimal value when the system is not decomposed into a binary and a single anymore, but by three bodies orbiting each other at the same SMA $a_{\rm max}$. From energy conservation, this condition is fulfilled when
\begin{equation}
\label{eq amax}
    \frac{a_{\rm max}}{a_0} = 1 + \frac{m_3m_{12}}{m_1m_2} = 1 + q_3 + \frac{q_3}{q_2}.
\end{equation}
In combination with equation (\ref{eq Nims basic}), we obtain
\begin{equation}
    \langle N_{\rm IMS}\rangle \simeq \left(   1 + q_3 + \frac{q_3}{q_2}     \right)^{\gamma-1}, \quad q_3<q_2.
\end{equation}
Because of symmetry we have 
\begin{equation}
\label{eq Nims theo}
   \langle N_{\rm IMS}\rangle \simeq \left(   1 + q_2 + \frac{q_2}{q_3}     \right)^{\gamma-1}, \quad q_3\ge q_2,
\end{equation}
because after the first intermediate state there is no memory of the initial conditions and the same arguments apply.
As for the exponent $\gamma$, in \cite{stone_leigh_2019} it is set at a value between 3.5 and 4, depending on the angular momentum of the system, while \cite{heggie_1975} finds 4.5\footnote{Note that these exponents are originally used for the distribution of the final energy, which is not necessarily the same as the distribution of the energy during one intermediate state, yet for this derivation we assume them to follow the same power-law.}.

The predictions from equation (\ref{eq Nims theo}) are displayed in Fig.~\ref{Nims theo}, where we choose $\gamma = 3.75$, so that $\langle N_{\rm IMS}\rangle\simeq 20$ in the equal-mass regime. We can see that, even though equation (\ref{eq Nims theo}) does not predict exactly the behaviour observed in the simulations, it can explain qualitatively why $\langle N_{\rm IMS}\rangle$ peaks if $q_2 = q_3$, and also it can explain the weaker dependence on $q_2\times q_3$. As it can be seen, the theoretical peak is narrower than the experimental one. This difference between theory and experiment can stem from the assumption that the temporary binary \textit{always} contains the two most massive bodies. It may be the case that the peak widens by relaxing this condition, and allowing the temporary binary to contain sometimes the lightest body. The distribution would also be broader for smaller values of $\gamma$.

\subsection{Eccentricity distribution}

   \begin{figure}
   \centering
   \includegraphics[width=\linewidth]{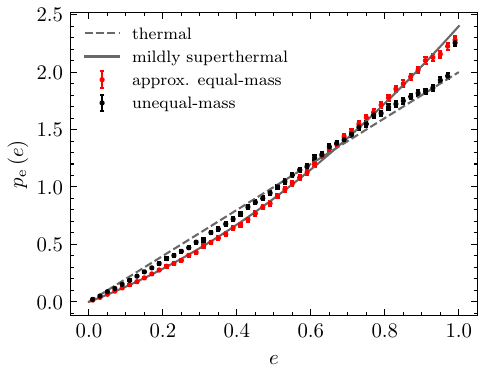}
      \caption{Eccentricity distribution after a democratic resonance. For the approximate equal mass case, the data points were selected by imposing the condition $q_2,\;q_3\in[0.5,1]$. For the unequal-mass case, the data points were selected by imposing the condition $q_2\in[0.1,0.4]$ or $q_3\in[0.1,0.4]$. The dashed lines represent the thermal distribution, and the solid line represents the mildly superthermal distribution.
              }
         \label{ecc}
   \end{figure}

It is often  assumed that the eccentricity of the binary that remains after a resonant encounter follows a thermal distribution $p_{e}\left(e\right)= 2e$.  \cite{stone_leigh_2019} show that for 
high angular momentum binary-single interactions\footnote{We simulate a range of impact parameters hence angular momenta, but the majority of encounters have high $b$ because of the uniform sampling in $b^2$. In the equal-mass limit and with $\langle e\rangle=2/3$, the angular momentum of the single BH is an order of magnitude higher of that of the binary. } the resulting eccentricity distribution is mildly super-thermal: $p_{e}\left(e\right) = (6/5) e\left(1+e\right)$.
In Fig.~\ref{ecc} we compare the eccentricity distribution in the equal and unequal-mass regimes with both the thermal and the mildly superthermal distributions. The eccentricity in the equal-mass case agrees with the mildly superthermal distribution, which agrees with the result of \cite{stone_leigh_2019}. Meanwhile, in the unequal-mass regime, the data agrees with a thermal distribution. According to \cite{ginat_2023}, an equal-mass binary interacting with a lower mass body is theoretically expected to have a highly superthermal distribution. However, our findings suggest that lower $m_3$ leads to a slight reduction in  the final eccentricities. We note that our sample contains unequal-mass binaries and a range of $b$, hence the results can not be directly compared.


Note that in the unequal-mass case, although the eccentricity agrees with a thermal distribution, there is a peak at $e\simeq1$. This peak is an artifact from our simulations that appears in encounters with $q_3 \simeq 1$, $q_2 \simeq 0.1$ and zero impact parameter. For these encounters, the secondary component of the binary is so light that it barely perturbs the other two bodies. Therefore, the secondary mass orbits a principal mass that is nearly sitting still. In the other hand, the single body, as massive as the primary mass and with zero impact parameter, follows a head-on collision towards the binary's centre of mass, almost where the primary mass rests. In this regime, the eventual interactions are extremely eccentric. This artifact appears because of the way that $b$ is sampled: we simulate a lot more encounters with impact parameter being \textit{exactly} zero than in nature, where the probability of a head-on encounter is zero. If we exclude the set of interactions with $b=0$, this artifact disappears.

\section{Application to globular cluster conditions}
\label{model}

In this Section we apply our results to the conditions in a globular cluster, which includes the properties of its BH population, its central BBH, and the conditions that lead the BBH into either a GW capture, an inspiral, or an ejection. Our model follows closely the model of  \cite{antonini_gieles_2020}.

\subsection{Globular cluster properties}
\label{cluster properties}
The total energy of the cluster is given by $E_{\rm cl} \simeq -0.2GM^2_{\rm cl}/r_{\rm h}$, with $M_{\rm cl}$ the cluster mass and $r_{\rm h}$ the half-mass radius. In steady post-collapse evolution, the  rate of change of $E$ is \citep{henon_1961,gieles_heggie_2011,breen_heggie_2013}
\begin{equation}
\label{eq Edot}
    \dot E_{\rm cl} = \zeta\frac{|E_{\rm cl}|}{t_{\rm rh}},
\end{equation}
with $\zeta \simeq 0.1$ \citep{henon_1961,alexander_gieles_2012,gieles_heggie_2011}, and $t_{\rm rh}$ the half-mass relaxation time, given by \citep{spitzer_hart_1971}
\begin{equation}
    t_{\rm rh} = 0.138\sqrt{\frac{M_{\rm cl}r^3_{\rm h}}{ G}}\frac{1}{\langle m_{\rm all}\rangle \psi \ln\Lambda}.
\end{equation}
Here, $\langle m_{\rm all}\rangle \simeq 0.5\, M_{\odot}$, $\ln\Lambda \simeq 10$, is the Coulomb logarithm which we assume to be constant, and $\psi=1$ for equal-mass clusters. The one-dimensional velocity dispersion of the cluster is
\begin{equation}
    \sigma \simeq \sqrt{\frac{{ G}M_{\rm cl}}{6r_{\rm h}}}.
\end{equation}
For a \citet{1966AJ.....71...64K} model with $W_0=7$, the central escape velocity is given by \citep{antonini_gieles_2020}
\begin{equation}
    v_{\rm esc} \simeq 35\, \kms\, \left(\frac{M_{\rm cl}}{10^5\,{\rm M}_{\odot}}\right)^{1/2} \left(\frac{1\,{\rm pc}}{ r_{\rm h}}\right)^{1/2} .
\end{equation}

\subsection{Properties of the BH population}
\label{model BH population}

We assume the presence of a single BBH at the core of the cluster, which contains the heaviest BH within the GC, with mass $m_1=m_{\rm max}$, and the second heaviest BH, with mass $m_2 \in[5\,{\rm M_{\odot}},m_1]$. In this range we assume a power-law mass function with slope $\alpha$ and adopt two values for $\alpha$ and $m_{\rm max}$ to mimic metal-poor and metal-rich progenitor stars. We use the rapid single-star evolution (SSE) code \citep{2000MNRAS.315..543H} with recent updates for massive star winds and BH masses by \citet{2020A&A...639A..41B} to determine BH mass functions for 0.01 solar ($Z=1.4\times10^{-4};\feh=-2$) and solar metallicity ($Z=0.014;\feh=0$). We find $m_{\rm max}\simeq25\,{\rm M}_{\odot}$ for metal-rich clusters, and $m_{\rm max}\simeq50\,{\rm M}_{\odot}$ for metal-poor clusters. Following \cite{heggie_1975, antonini_gieles_2023}, the BBH has a mass ratio $q_2$ that follows a distribution $p_2\left(q_2\right) \propto q_2^{\alpha_2}$, with $\alpha_2 = 3.5 +\alpha$. Here, $\alpha=-0.5$ for metal-poor clusters, and $\alpha=-2.3$ for metal-rich clusters. At the core we assume a density of field BHs $n_{\rm c}\propto m_3^{\alpha+1}$, where the additional $+1$ in the index (i.e., flatter) is because of mass segregation \citep{simon_2007}\footnote{We note that this flattening was not considered by \cite{antonini_gieles_2020}.}. The field BHs that interact with the binary have a  mass $m_3\in[5\,{\rm M_{\odot}},m_1]$ that follows a distribution $p_3\left(m_3\right) \propto n_{\rm c}(m_3) m_3^{1/2}\propto m_3^{\alpha_3} $, with $\alpha_3 = 3/2 +\alpha$. The additional slight flattening with respect to the central BH mass function is because of the assumption of equipartition, which (in the gravitational focussing limit) increases the interaction rate with more massive BHs because of their lower velocities \citep{antonini_gieles_2020}. In principle, it is possible that $m_3>m_1$. In fact, if we let $m_3$ in the same range as $m_1$, then around 15\%(5\%) of the times $m_3$ is larger than $m_1$ for metal-poor(metal-rich) clusters. Nevertheless, if the primary mass happened to not be the heaviest mass in the cluster, then the first interaction between the binary and that heaviest BH would most likely result in an exchange \citep{hills_fullerton_1980}, where the heaviest BH  becomes the new primary. We therefore only consider $m_3<m_1$.

Additionally, we assume that the principle of equipartition holds true for the BH population \citep{heggie_1975}. The principle states that the kinetic energy is the same for all BHs, implying that $\beta = (m_3\sigma_3^2)^{-1}$ is a constant of the BH population. This allows us to relate the one-dimensional velocity dispersion of one mass species $\sigma_3$, of mass $m_3$, with the one-dimensional velocity dispersion of the cluster 
\begin{equation}
\label{eq sigma3}
    \sigma_3 = \sigma\sqrt{\frac{m_{\rm ref}}{m_3}}, 
\end{equation}
where we set $m_{\rm ref}$ to be the mass species whose velocity dispersion is equal to the cluster's. This mass of reference is set at $10\,{\rm M}_{\odot}$.



\subsection{GW captures}
\label{GW captures}

During a resonant encounter two of the three BHs can merge if they pass sufficiently close such that GW emission causes a coalescence before the interaction would end. We refer to this as a GW capture. Following \cite{samsing_2014} we can  determine a posteriori what encounters undergo this outcome. A binary-single resonant encounter can be approximated as a sequence of $N_{\rm IMS}$ intermediate states composed of a temporary BBH with SMA $a_{\rm IMS}$ and eccentricity $e_{\rm IMS}$, and a single bound BH orbiting the binary in a wider orbit. We can define a critical distance $r_{\rm GW}$, below which the binary  merges due to GW energy loss before the return of the single BH. In \cite{samsing_2018}, $r_{\rm GW}$ is set to be the binary's pericenter at which a single passage radiates GW energy equal to the binding energy of the initial binary. The distance $r_{\rm GW}$ is then obtained from \cite{hansen_1972}:
\begin{equation}
    r_{\rm GW} \simeq 2.68 \left[ \frac{{ G}^{5/2}}{c^5} m_im_j\left(m_i+m_j\right)^{1/2} a \right]^{2/7},
\end{equation}
where $m_i$ and $m_j$ are the masses of the involved BHs. For a fixed SMA, only highly eccentric orbits have a small enough pericenter to trigger a GW capture. For this reason we have taken the limit $e\rightarrow 1$.

\subsection{In-cluster inspirals}
\label{inspirals}
In addition to GW captures, BBHs can also merge after a resonant encounter, but before the next one. We refer to this outcome as an inspiral. Throughout a binary's life inside the cluster,  successive interactions with single BHs or stars decrease the binary's SMA. This gradual hardening of the binary can lead to two different outcomes: either the binary is ejected from the cluster from the recoil it experiences after a resonant interaction, or the SMA is small enough that the BBH  inspirals before its next interaction with another single BH or star. The timescale between two successive resonant interactions, $\tau_{\rm r}$, can be obtained by noting that the binding energy increases on average by a fraction $\Delta_{\rm r}$, given in equation~(\ref{eq delta m3}). The timescale $\tau_{\rm r}$ is then obtained after the relation $\dot{\mathcal{E}} \simeq  \Delta_{\rm r} \mathcal{E}/\tau_{\rm r}$ \citep{heggie_hut_2003}. As we explain later on, the value of $\dot{\mathcal{E}}$ follows from the cluster relaxation process, and is set to be equal to $\dot E_{\rm cl}$. At later times, the evolution of the SMA is dominated by GW energy loss, described by \cite{peters_1964}
\begin{equation}
    \dot a_{\rm GW} = -\frac{64}{5}\frac{{G}^3m_im_j\left(m_i+m_j\right)}{{ c}^5a^3l^7}\left(1 + \frac{73}{24}e^2+\frac{37}{96}e^4\right),
\end{equation}
\begin{equation}
    \dot e_{\rm GW} = -\frac{304}{15}\frac{G^3m_im_j\left[m_i+m_j\right)}{{ c}^5a^4l^5}\left(e + \frac{121}{304}e^3\right),
\end{equation}
where  $l\equiv\sqrt{1-e^2}$ is the dimensionless angular momentum. The timescale of GW inspiral is then $\tau_{\rm GW} = a / |\dot a_{\rm GW}|$. In this model, the BBH inspirals if $\tau_{\rm GW} < \tau_{\rm r}$. Equivalently, this happens when the dimensionless angular momentum is lower than
\begin{equation}
\label{eq lgw}
    l < l_{\rm GW} \equiv \left[\frac{85}{3} \frac{G^4\left(m_im_j\right)^2\left(m_i+m_j\right) }{{ c}^5a^5\dot E_{\rm cl}}\Delta_{\rm r} \right]^{1/7},
\end{equation}
where we have taken the limit $e\rightarrow 1$. From this expression we define the critical value of the eccentricity $e_{\rm GW}$, above which the BBH inspirals.

\subsection{SMA of the BBH}
\label{sma of the binary}
The maximum SMA a BBH can have, denoted by $a_{\rm hs}$, corresponds to the hard-soft boundary: $\mathcal{E} = \langle\beta^{-1}\rangle =  m_{\rm ref}\sigma^2$ \citep{heggie_1975}, such that
\begin{equation}
    a_{\rm hs} = \frac{Gm_1m_2}{2m_{\rm ref}\sigma^2}.
\end{equation}
 We do not consider higher SMAs because binaries with $\mathcal{E}< m_{\rm ref}\sigma^2$ can be easily ionised by field BHs. Due to energy and momentum conservation, after the BBH interacts with a single BH, it experiences a recoil kick $v_{\rm kick}^2 \simeq \Delta_{\rm r} { G}{m_1m_2m_3}/(a{m_{123}m_{12}})$. When the binary's SMA is sufficiently low, the recoil kick can exceed the escape velocity of the cluster. The critical SMA at which ejection happens, denoted by $a_{\rm ej}$, is obtained imposing $v_{\rm kick} = v_{\rm esc}$ \citep{antonini_rasio_2016}:
\begin{equation}
\label{eq a_ej}
    a_{\rm ej} = \Delta_{\rm r} {G}\frac{m_1m_2m_3}{m_{123}m_{12}}\frac{1}{v^2_{\rm esc}}.
\end{equation}
For $a \le a_{\rm ej}$ and $m_3=\langle m_3\rangle$, the average three-body interaction ejects the binary from the cluster. The binary can also inspiral before it reaches $a_{\rm ej}$. If the BBH is initially assembled at the hard-soft boundary, in between each interaction there is a chance the BBH inspirals. Eventually, after many encounters, the probability of inspiral adds up to 1, and the sequence stops. We define the SMA at which this happens as $a_{\rm GW}$. Following \cite{antonini_gieles_2020}, the value of $a_{\rm GW}$ is obtained by imposing the condition $l_{\rm GW}^2(a_{\rm GW}) = 10/7\Delta_{\rm r}(1+\Delta_{\rm r})^{-1}$, which leads to
\begin{equation}
\label{eq agw}
    a_{\rm GW} =1.52\left(\frac{1+\Delta_{\rm r}}{\Delta_{\rm r}}\right)^{7/10}\left[ \frac{{G}^4(m_1m_2)^2m_{12}}{{c}^5\dot E_{\rm cl}}\Delta_{\rm r}  \right]^{1/5}
\end{equation}
A more detailed derivation can be found in the referenced paper. The value of the minimum SMA the BBH can attain, denoted by $a_{\rm m}$, is
\begin{equation}
    a_{\rm m} = {\rm max}\left(a_{\rm GW},a_{\rm ej}\right).
\end{equation}
Depending on the values of $a_{\rm ej}$ and $a_{\rm GW}$, the gradual hardening of the binary ends up either with an ejection, or with an inspiral.

In-cluster binaries then have a SMA $a\in[a_{\rm m},a_{\rm hs}]$. The exact distribution of $a$ is obtained following Hénon's principle \citep{henon_1975}. According to it, the flow of energy through the half-mass radius is independent of the precise mechanisms for energy production within the core, and the heat supplied to the cluster is assumed to be predominantly coming from the binding energy of the central BBH
\begin{equation}
\label{eq henon}
    \dot{\mathcal{E}} =  \dot E_{\rm cl},
\end{equation}
with $E_{\rm cl}$ the energy of the cluster, described in Section \ref{cluster properties}. Note that both energies have equal signs because $\mathcal{E}$ is defined in equation (\ref{bin}) as the absolute value of the binding energy. Because $\dot E_{\rm cl}$ is constant \citep{henon_1975},  it follows that
\begin{equation}
\label{eq SMA}
    p_{a}\left(a\right) = \frac{a_{\rm hs}a_{\rm m}}{a_{\rm hs}-a_{\rm m}}  \frac{1}{a^2}.
\end{equation}
In practise only one BBH is present in a cluster, such that $p_{a}(a)$ represents the probability for the BBH to have a SMA in the range $[a_{\rm m}, a_{\rm hs}]$.

\section{Results: implications for GWs}
\label{results 2}

In this Section we combine the GC model of Section~\ref{model}  with the binary-single simulations of Section~\ref{results 1} to study the GW implications of unequal-mass BH encounters. From now on, the bodies that constitute our simulated binary-single scatterings are going to be referred as BHs.

\subsection{Probability of a GW capture}

   \begin{figure*}
   \includegraphics[width=0.5\linewidth]{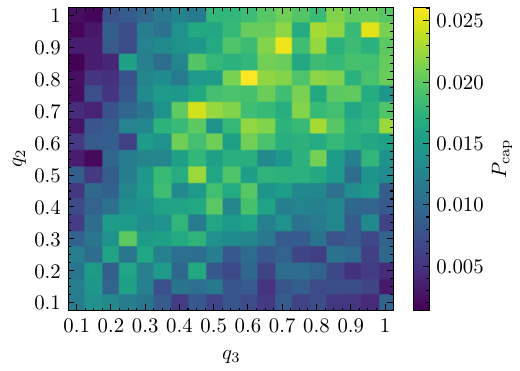}
    \includegraphics[width=0.5\linewidth]{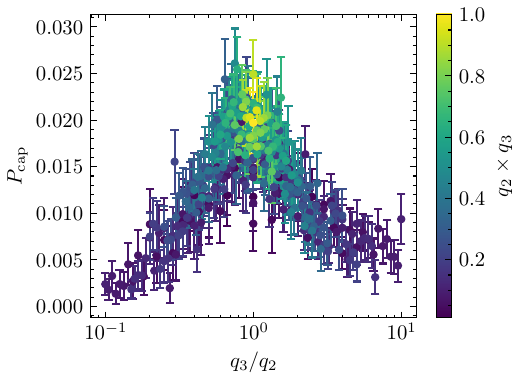}
              

  
      \caption{{\it Left:} $P_{\rm cap}$ as a function of $q_2$ and $q_3$.
             {\it Right:} $P_{\rm cap}$ as a function of $q_2/q_3$. For that, we used $m_1=45{\rm M}_{\odot}$, $a=a_{\rm GW}$, and assumed a metal-poor cluster with $M_{\rm cl} = 5\times10^5\,{\rm M}_{\odot}$ and $r_{\rm h} = 1\,{\rm pc}$. The error bars are computed via $\sqrt{2N_{\rm cap}}/N_{\rm bin}$, with $N_{\rm bin}$ the number of resonant events in the bin, and $N_{\rm cap}$ the number of GW captures among them. 
              }
         \label{pcap}
   \end{figure*}

According to \cite{samsing_2018}, the probability that a binary-single resonant interaction ends with a GW capture (denoted by $P_{\rm cap}$) is approximated by 
\begin{equation}
\label{eq pcap}
    P_{\rm cap} \simeq \langle N_{\rm IMS}\rangle \frac{2r_{\rm GW}}{a}.
\end{equation}

In Fig.~\ref{pcap} we present the observed $P_{\rm cap}$ from our scattering experiments, as a function of the mass ratios $q_2$ and $q_3$. From this we see that $P_{\rm cap}$ is maximum  if $q_2\simeq q_3$. This is because then $\langle N_{\rm IMS}\rangle$ is highest (Section \ref{subsection nims}).



\subsection{Differential rate of mergers}
\label{section differential rate}

   \begin{figure}
   \centering
   \includegraphics[width=\linewidth]{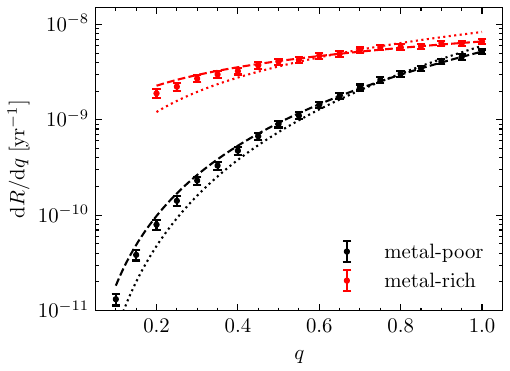}
      \caption{Mass-ratio distribution of in-cluster mergers, for a metal-poor (black) and metal-rich (red) clusters with $M_{\rm cl} = 2\times10^5\,{\rm M}_{\odot}$ and $r_{\rm h} = 3 \,{\rm pc}$. The dashed lines represent the theoretical slopes given in equation (\ref{eq Rq}) and the dotted lines represent $p_2(q_2)$. Both the dashed and dotted lines have been normalised to the same normalisation constant as the data points.
              }
         \label{fig Rq}
   \end{figure}


A GC can be characterised by its $M_{\rm cl}$, $r_{\rm h}$ and metallicity. 
The metallicy affects mostly the properties of the BH mass function, while $M_{\rm cl}$ and $r_{\rm h}$ determine the GW signature.
The mass ratio of merging BBHs, $q$, also depends on the GC properties. In this Section we explore how the merger rate, $R$, depends on the mass ratio $q$, that is, ${\rm d}R/{\rm d}q$, and we derive a theoretical prediction for it.

Binary-single interactions can lead to mergers between BHs 1-2, 1-3 and 2-3, as the result of both GW captures and inspirals in between resonant interactions (hereafter referred to as inspirals).  As we will later see, the majority of mergers come from inspirals between the BHs that constitute the initial binary (with mass ratio $q=q_2$). Hence, for the purpose of this derivation, we approximate all mergers to come from this channel. 

Under the assumption that there is a single BBH at any time, the differential merger rate can be written as
\begin{equation}
\frac{{\rm d}R}{{\rm d}q} = \Gamma_{\rm b}(m_1,q)p_2(q),
\end{equation}
where $\Gamma_{\rm b}(m_1,q)$ is the mass-dependent binary formation rate and $p_2(q)$ is the probability a dynamical binary has mass ratio $q$ (see Section \ref{model BH population}). $\Gamma_{\rm b}(m_1,q)$ is the inverse of the time that it takes for the BBH to harden from the hard-soft boundary to $a_{\rm GW}$. From Hénon's principle it follows that
\begin{equation}
    \dot E_{\rm cl} = \frac{{\rm d}}{{\rm d}t}\left(\frac{{ G}m_1m_2}{2a}\right).
\end{equation}
By integrating this differential equation from $a_{\rm hs}$ to $a_{\rm GW}$ we obtain
\begin{equation}
    \Gamma_{\rm b}(m_1,q) \simeq \frac{2a_{\rm GW}\dot E_{\rm cl}}{{ G}m_1m_2},
\end{equation}
where we assumed $a_{\rm GW}\ll a_{\rm hs}$ and the explicit value of $a_{\rm GW}$ is given in equation (\ref{eq agw}) such that 
\begin{equation}
\label{eq delta t}
    \Gamma_{\rm b}(m_1,q)\propto \frac{1}{m_1}\frac{\left(1+q\right)^{1/5}}{q^{3/5}}.
\end{equation}
Consequently
\begin{equation}
\label{eq Rq}    \frac{{\rm d}R}{{\rm d}q} \propto \frac{1}{m_1}\frac{\left(1+q\right)^{1/5}}{q^{3/5}} p_2(q).
\end{equation}
Note that the previous equation can be approximated by
\begin{equation}
    \frac{{\rm d}R}{{\rm d}q}\propto \frac{1}{m_1}q^{2.9+\alpha},
\end{equation}
with $\alpha$ the exponent of the BH mass function. Because $M_{\rm cl}$ and $r_{\rm h}$ do not enter into equation~(\ref{eq Rq}), the shape of ${\rm d}R/{\rm d}q$ is entirely determined by the shape of the BH mass function, which is a function of metallicity and the stellar initial mass function. This means that the observed ${\rm d}R/{\rm d}q$ of dynamically formed BBH can be used to infer the underlying BH mass function.

We now compare the result from equation (\ref{eq Rq}) to the differential rates that were found combining our scattering events with our GC model. The exact methodology used to obtain ${\rm d}R/{\rm d}q$ from the scattering experiments is described in Appendix~\ref{APPENDIX f(q)}. Contrary to equation (\ref{eq Rq}), here we take mergers from both GW captures and inspirals into account, involving BHs 1-2, 1-3 and 2-3. We can see in Fig.~\ref{fig Rq} the results for metal-poor and metal-rich clusters, along with their respective predictions from equation (\ref{eq Rq}). The results from the scattering experiments for both metallicities agree with the result of equation (\ref{eq Rq}). We previously assumed that the majority of mergers involve the two BHs that were in the initial binary. Here we quantify this: approximately 70\% of mergers are between BH 1 and 2; 20\% are between BH 1 and 3 and mergers between the BH 2 and BH 3 contribute  around 10\%. 

As seen in Fig.~\ref{fig Rq}, for equal $M_{\rm cl}$ and $r_{\rm h}$, metal-rich clusters exhibit a higher rate of mergers than metal-poor clusters, by about a factor 2. This is to be expected from Hénon's principle. A cluster requires its central BBH to provide a fixed quantity of energy per unit of time, independently of the characteristics of the BBH. A metal-rich cluster hosts a less massive central BBH, and to keep up with the energy demand of the cluster, the central BBH needs to shrink its orbit at a faster rate, leading to a quicker coalescence. This is explicitly seen in equation (\ref{eq delta t}) ($\Gamma_{\rm b} \propto m_1^{-1}$). We adopted a primary mass for the metal-rich cluster that was half that of the metal-poor clusters, increasing the rate. In addition, the steeper BH mass function at higher metallicity ($\alpha=-2.3$ vs. $\alpha=-0.5$) results in binaries with lower $q$, preferentially increasing the rates at low $q$. 


\subsection{Rate of mergers}
By integrating the differential rates shown in Fig.~\ref{fig Rq}, we  obtain the total rates that are expected for clusters with $M_{\rm cl} = 2\times10^5\,{\rm M}_{\odot}$ and $r_{\rm h} = 3 \,{\rm pc}$. This results in $R_{\rm mr} = 3.7\,{\rm Gyr}^{-1}$, and $R_{\rm mp} = 1.5\,{\rm Gyr}^{-1}$, for the metal-rich and metal-poor cluster, respectively.


These rates were computed using a distribution of $m_2$ and $m_3$. To see how much things change by including unequal-mass BHs with respect to the equal-mass case, we also compute these rates assuming all interactions to happen among BHs of equal masses. That is, we use $p_2(q_{2})=\delta(1-q_{2})$ and $p_3(q_{3})=\delta(1-q_{3})$ (therefore, all masses involved in the three-body problem are equal to the primary mass). The resulting equal-mass rates are $R_{\rm EM, mr} = 3.5\times10^{-9}\,{\rm yr}^{-1}$ and $R_{\rm EM, mp} = 1.5\times10^{-9}\,{\rm yr}^{-1}$. These values are within a 5\% of the unequal-mass rates shown above, which implies that the equal-mass assumption is a good approximation of the more realistic unequal-mass case. Nonetheless, this approximation is good as long as the involved masses are all equal to the primary mass, which is considered the highest mass in the cluster.


We now present the rate of mergers that is expected from a Milky Way-like GC population. We sample $M_{\rm cl}$ and $r_{\rm h}$ from the log-normal distributions
\begin{equation}
    \phi_{\rm M}\left(M_{\rm cl}\right)\propto \frac{1}{M_{\rm cl}}{\rm exp}\left(-\frac{{\rm log}_{10}^2\left(M_{\rm cl}/\mu_{M}\right)   }{ 2\sigma_{M}^2 }\right),
\end{equation}
\begin{equation}
    \phi_{\rho}\left(\rho_{\rm h}\right)\propto \frac{1}{\rho_{\rm h}}{\rm exp}\left(-\frac{{\rm log}_{10}^2\left(\rho_{\rm h}/\mu_{\rho}\right)   }{ 2\sigma_{\rho}^2 }\right),
\end{equation}
with $\rho_{\rm h}\equiv 3M_{\rm cl}/(8\pi r_{\rm h}^3)$ the half-mass density, $\mu_{\rho}=5\times10^2\,{\rm M}_{\odot}{\rm pc}^{-3}$, $\sigma_{\rho}=0.75$, $\mu_{M}=2\times10^5\,{\rm M}_{\odot}$, $\sigma_{M}=0.5$, and with the limits $M_{\rm cl}\in[10^2,\,10^7]\,{\rm M}_{\odot}$ and $\rho_{\rm h}>0$. For a number density of GCs in the Universe of $4\times10^9\,{\rm Gpc^{-3}}$ \citep{antonini_gieles_2020b}, we obtain
\begin{equation}
    R_{\rm mr} = 4.4^{+1.6}_{-1.2} \times\frac{f_{\rm mr}}{0.5}\,{\rm Gpc^{-3}yr^{-1}},
\end{equation}
\begin{equation}
    R_{\rm mp} = 2.0^{+0.8}_{-0.4}\times\frac{1-f_{\rm mr}}{0.5} \,{\rm Gpc^{-3}yr^{-1}},
\end{equation}
where $R_{\rm mr}$ ($R_{\rm mp}$) is the contribution from metal-rich (metal-poor) GCs, and $f_{\rm mr}$ is the fraction of metal-rich clusters. For this computation, we sampled 1000 different clusters, and to save computation time we used the equal-mass approximation.

\subsection{Inspirals driven by direct encounters}

   \begin{figure}
   \centering
   \includegraphics[width=\linewidth]{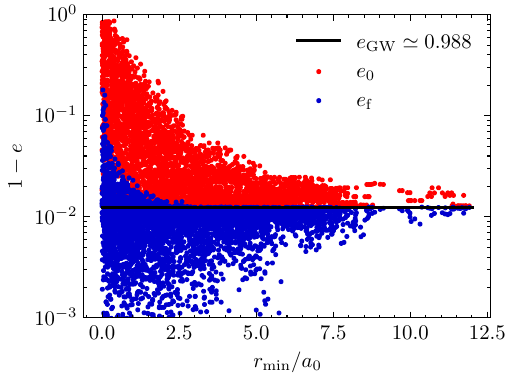}
      \caption{ Eccentricity of the subset of binaries that undergo direct inspiral, as a function of the minimum distance at which the field BH approaches one of the two components of the binary $r_{\rm min}$. In red, the initial eccentricity, and in blue, the final eccentricity. The solid line represents the critical eccentricity $e_{\rm GW}$, calculated with  $M_{\rm cl} = 5\times10^5{\rm M}_{\odot}$, $r_{\rm h} = 1 {\rm pc}$ and $\alpha = -0.5$. The binary has equal-mass components with $m_{12}=40\,{\rm M}_{\odot}$, a SMA of 0.5 AU, and the field BH has a mass in the range $m_3\in[0.1,0.5]m_{12}$.
              }
         \label{ins_ecc}
   \end{figure}

As explained in Section \ref{model}, inspirals happen when the binary has an eccentricity sufficiently high so that the timescale of merger is shorter than the timescale between resonant interactions. As considered in \cite{antonini_gieles_2020}, these high eccentricities can be achieved when, after a resonant interaction, the post-encounter eccentricity has a value above $e_{\rm GW}$. Nonetheless, in our simulations only around 60\% of the inspiraled BBHs achieved their high $e$ through this channel. The remaining 40\% achieved it after a direct (i.e., non-resonant) encounter. As described in \cite{heggie_rasio_1996}, direct encounters can modify the binary's eccentricity, both positively and negatively, with bigger changes becoming more frequent in closer approaches. For a binary that is just below the critical eccentricity, this can cause a positive change in its eccentricity, potentially pushing it into an inspiral. This mechanism is referred to as a direct inspiral. Given how frequent direct encounters are in comparison to resonances, these two channels end up producing approximately the same number of inspirals.

Fig.~\ref{ins_ecc} represents the initial ($e_0$) and final ($e_{\rm f}$) eccentricities of the subset of binaries from our scattering experiments that undergo a direct inspiral. These inspirals can happen for field BHs that approach the binary with a distance of up to $\sim10 a_0$, but the further away the approach, the closer $e_0$ has to be to $e_{\rm GW}$ in order for the binary to inspiral. For distances below $\sim3a_0$, some inspirals happen with $e_{\rm f}$ above the solid line. For these close encounters, the binary's SMA decreases after the encounter, changing $e_{\rm GW}$ to lower values. For this subset of close encounters, what drives the inspiral is not only the increase in eccentricity, but also the decrease of the SMA. In  Appendix~\ref{APPENDIX sigma}, we derive an approximation of the cross section of direct inspirals ($\Sigma_{\rm d}^{\rm ins}$), and we compare it to the cross section of inspirals coming from resonances ($\Sigma_{\rm r}^{\rm ins}$). We find a ratio of $\Sigma_{\rm d}^{\rm ins} / \Sigma_{\rm r}^{\rm ins} \simeq 0.84$, reasonably close to the experimental ratio of around 2/3.

Most direct inspirals were found to happen in interactions with an inclination of $\theta\sim \pi/2$. This is expected from  equation (15) of \cite{heggie_rasio_1996}, which states that $\delta e \propto {\rm sin}^2\theta$, therefore maximising the changes in eccentricity when $\theta= \pi/2$. In addition, direct inspirals may have a different eccentricity signature than inspirals coming from resonant interactions.

Direct inspirals are not included in fast models for dynamical production of BBH inspirals, such as \textsc{cBHBd} \citep{antonini_gieles_2023}. There is currently a gap in the prediction of in-cluster mergers between these fast models and more accurate $N$-body models \citep{rastello_2019,kremer_2020,banerjee_2021,barber_2023}, where the latter predicts a higher proportion of BBHs coalescing inside the cluster instead of after ejection. \cite{pina_2023} suggest that the aforementioned fast models  lack an important ingredient which may compensate for the lack of in-cluster mergers, namely interactions between the dynamically active BBH and wide BBHs that form dynamically at a high rate. Direct inspirals provides an additional ingredient to bridge the gap between fast models and $N$-body simulations.

So far, we have only considered BHs as the perturbers that trigger these direct inspirals. As it can be seen in equation (\ref{eq sigma ecc}), $\Sigma_{\rm d}^{\rm ins} \propto m_3^{5/3}$. This implies that the cross section for direct inspiral decreases rapidly for lower mass perturbers. Namely, for stars, $m_3$ decreases by around a factor $\gtrsim20$, from $\langle m_3\rangle \simeq 10 \,{\rm M}_{\odot}$ for BHs, to $\langle m_3\rangle\simeq0.5\, {\rm M}_{\odot}$ for stars, which implies that $\Sigma_{\rm d}^{\rm ins}$ decreases by a factor $\gtrsim150\,$. In order for stars to become relevant perturbers, their number density should surpass that of BHs by around that factor, which is very unlikely in the core of a GC.

According to \cite{heggie_rasio_1996}, the cross section for a change in eccentricity $\delta e$ is symmetric with respect to the sign of $\delta e$. This symmetry implies that both eccentricity excitation and de-excitation are equally possible. Note that, so far, we have focused on the possibility of BBHs inspiraling due to an eccentricity increase through a direct encounter, but there is also the possibility of the opposite to happen. What if, through eccentricity de-excitation, direct encounters \textit{prevent} BBHs from coalescing? 

Here we argue why we believe this is unlikely. Let us consider a BBH in the process of inspiraling (that is, with $e>e_{\rm GW}$). By definition, this binary fulfills $\tau_{\rm GW} < \tau_{\rm r}$, with $\tau_{\rm GW}$ the upper limit of the remaining lifetime of the BBH, and $\tau_{\rm r}$ the timescale between resonant interactions. Using again the symmetry argument in the sign of $\delta e$, we approximate the cross section of a direct inspiral (with timescale $\tau_{\rm d}^{\rm ins}$) to be the same as the cross section of a direct encounter to decrease the eccentricity of a BBH with $e>e_{\rm GW}$ to lower values (a coalescence prevention, with timescale $\tau_{\rm d}$). Therefore, $\tau_{\rm d}\simeq\tau_{\rm d}^{\rm ins}$. As mentioned previously, we find in our scattering experiments that $\Sigma_{\rm d}^{\rm ins}\lesssim\Sigma_{\rm r}^{\rm ins}$, hence $\tau_{\rm r}^{\rm ins}\lesssim\tau_{\rm d}^{\rm ins}$. In this last step we just assumed that timescales and cross sections are inversely proportional. Moreover, mergers caused by resonant interactions are only a subset of all resonant interactions. Consequently, $\tau_{\rm r}$ is shorter than $\tau_{\rm r}^{\rm ins}$. Summarizing, we have
\begin{equation}
    \tau_{\rm GW} < \tau_{\rm r} < \tau_{\rm r}^{\rm ins} \lesssim \tau_{\rm d}^{\rm ins}\simeq \tau_{\rm d}.
\end{equation}
In other words, $\tau_{\rm GW}<\tau_{\rm d}$ implies that, by the time a direct encounter arrives to prevent the coalescence, it is already too late.

\section{Discussion}
\label{sec discussion}



Our model is similar to that of \citet{antonini_gieles_2020}, and we add the variations of the two mass ratios that are involved in binary-single interactions.
In this section we describe the ingredients that are missing in our analysis.

\subsection{Missing ingredients}

We did not include  evolutionary effects such as GC mass loss and the dynamical ejection of BH. This means that the BH mass function is always fully populated. This biases our mergers to low $q$, because in an evolving GC, the most massive BHs gradually get ejected from the GC and are therefore no longer available for mergers. This is more important for metal-rich GCs that start with a lower $m_1$ and eject BHs at a faster rate.
Moreover, we assumed the presence of a single BBH in the core of our GC that only interacts with single BHs. The assumption of a single BBH is supported by results of $N$-body simulations \citep{pina_2023}, but the authors also show that there is a high rate of interactions between the hard BBH and shorted-lived (soft) BBHs. In addition, a large fraction of primordial massive binaries is expected to turn into BBHs, that can interacts with the dynamical BBHs \citep{2024MNRAS.527.7363B}. Although less frequent, binary-binary interactions are estimated to contribute 25\% to 45\% of all highly eccentric mergers \citep{zevin_2019}. Including them may change our analysis in a way that has not been explored.

Additionally, we only consider first generation mergers, that is, we do not consider mergers where at least one of the BHs has merged previously, referred to as hierarchical mergers. According to \cite{ye_2024}, hierarchical mergers approximately account for $\sim20\%$ of mergers and may populate the pair-instability mass gap \citep{antonini_gieles_2023}. In order to account for hierarchical mergers, GW kicks must be taken into account. Due to linear momentum conservation, the post-merger BH receives a GW kick that, depending on $q$ and the BH spins, can go from zero for non-spinning BHs with $q=1$, to several hundreds-thousands of km/s, which is enough to eject the post-merger BH from the cluster \citep{antonini_rasio_2016}. Therefore, we should expect hierarchical mergers to have a higher relevance in clusters with a high $v_{\rm esc}$, where post-merger BHs are more likely to be retained in the cluster. It is for this reason that we expect the results from Section \ref{section differential rate} to be more accurate for GCs of low $v_{\rm esc}$. In clusters where hierarchical mergers are indeed relevant, the post-merger BH, with mass $\lesssim m_1+m_2$, eventually falls back to the cluster's core and forms a new binary with the second heaviest object in the cluster. Consequently, this second generation binary is likely to have a lower mass ratio than the previous first generation binary. As seen in \cite{antonini_gieles_2023}, including hierarchical mergers increases the number of mergers at lower mass ratios.

\subsection{Newtonian dynamics}

The simulations were carried out using Newtonian dynamics, which allows us to determine the dynamics of the system not by the absolute masses, but rather by the mass ratios. This reduces the dimension of the parameter space by one and simplifies the analysis, but it does so at the cost of accuracy during close encounters \citep{gultekin_2006}. Therefore, our method to determine GW captures is only a rough approximation. Nevertheless, in-cluster coalescences are dominated by inspirals \citep{rodriguez_2018,antonini_gieles_2020}, which in our model are predicted not by close approaches during resonant encounters, but by the semi-major axes and eccentricities attained after them, which are less sensitive to post-Newtonian terms. Therefore, the lack of post-Newtonian terms may affect primarily GW captures, which are subdominant with respect to inspirals. In addition, in Appendix \ref{APPENDIX 1pn} we find $1^{\rm st}$ order post-Newtonian effects are safe to ignore in the specific case of direct inspirals.

\section{Conclusions}
\label{conclusions}

In this study we have investigated unequal-mass binary-single interactions in the Newtonian point-mass limit. For that, we have simulated a total of $\mathcal{O}(10^7)$ scattering events using \textsc{starlab}. Our key findings are:
\begin{enumerate}
    \item 
    The fractional change of binding energy following a resonant interaction, $\Delta_{\rm r}$, can be well described by $\Delta_{\rm r}=0.2\left[ 1-{\rm exp}(-Am_3/m_{12})\right]$ with $A\simeq7.0$ for $m_3<m_{12}$; 
    \item The number of intermediate states  peaks at  $N_{\rm IMS}\simeq20$ if the two lightest bodies in the triplet have similar masses ($m_2\simeq m_3$). Outside of this regime, $N_{\rm IMS}$ drops considerably, reducing the probability for GW captures;
    \item In agreement with \cite{stone_leigh_2019}, we found a mildly superthermal distribution of eccentricities near the equal-mass case with high angular momentum. Outside of the equal-mass case and considering lower $m_3$, the eccentricity distribution after resonant interactions is closer to thermal.
\end{enumerate}

We used our scattering experiments in combination with a model for the evolution of a BBH in a GC to explore the GW implications of unequal-mass encounters of BHs in GCs. We summarise the main results below.
\begin{enumerate}
   \item The differential merger rate for different mass ratios is ${\rm d}R/{\rm d}q\propto m_1^{-1}q^{-3/5}(1+q)^{1/5}p_2(q)$, with $p_2(q)$ the mass-ratio distribution of dynamically formed BBHs, which depends on the underlying BH mass function. This means that the mass-ratio distribution of mergers is slightly flatter than the mass-ratio distribution of the dynamical binaries;
    \item In equal conditions of cluster mass and radius, metal-rich clusters have approximately double the merger rate as metal-poor clusters;
    \item We compared the rate of mergers obtained assuming both the equal and the unequal-mass regimes, and found that the equal-mass case can approximate within a 5\% error the rates predicted by the latter, as long as the chosen masses are equal to the primary mass;
    \item While BBHs inspiral when their eccentricity is above a threshold $e_{\rm GW}(a)$ (according to our model), only 60\% of the inspiraled BBHs were found to attain their high eccentricities through resonant interactions. The other 40\% obtained their high eccentricity after a direct encounter. This mechanism was denominated as direct inspirals, and its contribution to mergers may be the missing ingredient that bridges the current gap between fast models and more accurate $N$-body models.
\end{enumerate}


\begin{acknowledgements}
BRF, DMP and MG  acknowledge financial support from the grants PID2021-125485NB-C22, EUR2020-112157, CEX2019-000918-M funded by MCIN/AEI/10.13039/501100011033 (State Agency for Research of the Spanish Ministry of Science and Innovation) and SGR-2021-01069 (AGAUR). BRF thanks the Leiden Observatory, where part of this work was done,  for their hospitality during a three months research stay.
\end{acknowledgements}

%
   \bibliographystyle{aa} 
   \bibliography{total} 
%


\begin{appendix}

\section{Derivation of ${\rm d}R/{\rm d}q$ from the scattering experiments}
\label{APPENDIX f(q)}

Here we show how we obtain the differential merger rate, ${\rm d}R/{\rm d}q$, for a GC with arbitrary $M_{\rm cl}$, $r_{\rm h}$ and metallicity based on the results of the gravitational scattering experiments presented in Section~\ref{results 1}. According to Hénon's principle, the rate at which the central BBH releases energy is determined by the cluster, and is independent of the BBH's properties. Assuming the binary only releases its energy through resonant binary-single interactions, this rate can be expressed as (see Section \ref{inspirals})
\begin{equation}
    \dot{\mathcal{E}} = \frac{\langle\Delta_{\rm r}\rangle \mathcal{E}}{\tau_{\rm r}}.
\end{equation}
We define $R_{\rm r}\equiv 1/\tau_{\rm r}$ as the rate at which the BBH undergoes resonant interactions. The rate at which BHs $i$ and $j$ coalesce, denoted by $R_{ij}$, is $R_{\rm r}$ times the number of mergers per resonant interaction, which is obtained via
\begin{equation}
\label{eq Rij1}
    R_{ij} = R_{\rm r}\frac{\Sigma_{ij}}{\Sigma_{\rm r}},
\end{equation}
where $\Sigma_{ij}$ and $\Sigma_{\rm r}$ are the cross sections for $i-j$ mergers and resonant interactions respectively. We find $\langle \Delta_{\rm r}\rangle$ from
\begin{equation}
    \langle\Delta_{\rm r}\rangle = \int_{m_{\rm min}}^{m_{\rm max}}\Delta_0\left[1- {\rm exp}\left(-A\frac{m_3}{m_{12}}\right)\right]p_3(m_3) {\rm d}m_3,
    \label{avD}
\end{equation}
where we use the fitting function given in equation (\ref{eq delta m3}), and $p_3(m_3)$ is given in Section \ref{model BH population}. We assume a preexisting BBH with mass ratio $q_2$, and an interaction with a single BH of mass ratio $q_3$. It is for this reason that equation~(\ref{eq Rij1}) must be weighed by $p_2(q_2)$ and $p_3(q_3)$, given in Section \ref{model BH population}. Therefore
\begin{equation}
\label{eq Rij}
    \frac{{\rm d}^2R_{ij}}{{\rm d}{q_2}{\rm d}{q_3}} = R_{\rm r}(q_2)\frac{\Sigma_{ij}}{\Sigma_{\rm r}}(q_2,q_3) p_2(q_2)p_3(q_3).
\end{equation}
The ratio $\Sigma_{ij}/\Sigma_{\rm r}(q_2,q_3)$ is obtained from our binary-single scattering events, which are described in Section \ref{STARLAB}. In our simulations we use values of $q_2$ and $q_3$ in the range $[0.1,1]$ and spaced in steps of 0.05. We perform the following steps:
\begin{enumerate}
    \item Determine the primary mass (and mass scale) $m_1$, equal to $25\,{\rm M}_{\odot}$ for metal-rich clusters, and $50\,{\rm M}_{\odot}$ for metal-poor clusters (see Section \ref{model BH population}),
    \item Select one of the 19 values of $q_2$ that were used in the simulations and determine the secondary mass $m_2 = m_1 q_2$, and select one of the 19 values of $q_3$ that were used in the simulations and determine the mass of the field BH $m_3=m_1 q_3$,
    \item For all encounters with this combination of $q_2$ and $q_3$, read $\Sigma_{\rm r}$.
    \item Sample the SMA of the binary from $p_{a}(a)$, with $a\in[a_{\rm m},a_{\rm hs}]$ (see Section \ref{sma of the binary}),
    \item For all encounters with this combination of $q_2$, $q_3$ and $a$, read $\Sigma_{ij}$, including both GW captures and inspirals. A GW capture between BHs $i$ and $j$ happens if the minimum distance between them that is reached during the whole encounter is below $r_{\rm GW}$ (see Section \ref{GW captures}). An inspiral between BHs $i$ and $j$ happens if the final BBH is composed by these two BHs, and has an eccentricity after the interaction that is above $e_{\rm GW}$ (see Section \ref{inspirals}). 
\end{enumerate}
Steps 4 and 5 are repeated for 10 randomly sampled SMAs, obtaining the ratio $\Sigma_{ij}/\Sigma_{\rm r}$ averaged over $a$, and steps 2 to 5 are repeated for all values of $q_2$ and $q_3$, thus obtaining $\Sigma_{ij}/\Sigma_{\rm r}(q_2,q_3)$. We can now use equation (\ref{eq Rij}) to determine ${\rm d}R/{\rm d}q$. Let us first consider mergers among BHs 1 and 2. For these mergers, $q=q_2$. In order to obtain ${\rm d}R_{12}/{\rm d}q$, we have to integrate over all possible $q_3$'s (in other words, we have to integrate along lines of constant $q$)
\begin{equation}
    \frac{{\rm d}R_{12}}{{\rm d}q} = \int_{q_{\rm min}}^1 \frac{{\rm d}^2R_{12}}{{\rm d}{q}{\rm d}{q_3}}{\rm d}q_3,
\end{equation}
Similarly, for mergers between BHs 1 and 3, we have $q=q_3$, and for mergers between BHs 2 and 3, we have $q={\rm min}(q_2,q_3)/{\rm max}(q_2,q_3)$. To obtain their respective differential rates, we integrate along lines of constant $q$, as in the previous equation. The differential rate that includes all types of mergers is
\begin{equation}
    \frac{{\rm d}R}{{\rm d}q} =  \frac{{\rm d}R_{12}}{{\rm d}q} + \frac{{\rm d}R_{13}}{{\rm d}q} + \frac{{\rm d}R_{23}}{{\rm d}q}.
\end{equation}
The results of this exercise are shown in Fig.~\ref{fig Rq}.

\section{Derivation of $\Sigma_{\rm d}^{\rm ins}$}
\label{APPENDIX sigma}

In this appendix we compute the ratio between the cross section of a direct inspiral $\Sigma^{\rm ins}_{\rm d}$ and that of a resonant inspiral $\Sigma^{\rm ins}_{\rm r}$. First, we estimate $\Sigma^{\rm ins}_{\rm r}$ with
\begin{equation}
    \Sigma^{\rm ins}_{\rm r} = \Sigma_{\rm r} P_{\rm ins},
\end{equation}
where $P_{\rm ins} = 1 - e_{\rm GW}^2$, and $\Sigma_{\rm r}$ is the cross section of resonance
\begin{equation}
    \Sigma_{\rm r} = A\pi a^2\tilde{v}^{-2},
\end{equation}
where $A$ is set to a value of 19.55 by integrating equation (24) from \cite{heggie_hut_1993}. For a binary of initial eccentricity $e$, the cross section for a direct encounter to change the eccentricity by a value $\delta e > \delta e_0$ is given by \cite{heggie_rasio_1996}
\begin{equation}
\label{eq sigma ecc}
    \Sigma\left(\delta e > \delta e_0\right) \simeq 4.29 \left(\frac{m_3^2}{m_{12}m_{123}}\right)^{1/3} \frac{m_3m_{12}a^2}{m_1m_2\tilde{v}^2} e^{2/3}\left(1-e^2\right)^{1/3} \delta e_0^{-2/3}.
\end{equation}
This expression assumes $r_{\rm p} \gg a$. Although in Fig.~\ref{ins_ecc} it can be seen that a big proportion of inspirals happen at $r_{\rm p} \simeq a$, our aim is not to obtain an exact value of $\Sigma^{\rm ins}_{\rm d}$, but rather to showcase that it is of the same order of magnitude than $\Sigma^{\rm ins}_{\rm r}$. It is for this reason that, even though the assumption $r_{\rm p} \gg a$ is false, we still use equation (\ref{eq sigma ecc}), assuming that it approximates the order of magnitude. 

To know the cross section for $\delta e$ to be high enough to trigger an inspiral, we set $\delta e_0 = e_{\rm GW} - e$. Now, we set $\Sigma^{\rm ins}_{\rm d}$ to be the expected value of $\Sigma\left(\delta e > \delta e_0\right)$ over all possible initial eccentricities:
\begin{equation}
    \Sigma^{\rm ins}_{\rm d} = \int_0^{e_{\rm GW}} p_{e}\left(e\right) \Sigma\left(\delta e > e_{\rm GW} - e\right) {\rm d}e,
\end{equation}
where we have assumed $p_{e}(e)$ to be a thermal distribution with $e$ in the range $e\in[0,\,e_{\rm GW}]$. For equal masses, the ratio of cross sections reduces to
\begin{equation}
\label{eq sigma ratio}
    \frac{\Sigma^{\rm ins}_{\rm d}}{\Sigma^{\rm ins}_{\rm r}} \simeq 0.15\frac{I(e_{\rm GW})}{1-e_{\rm GW}^{2}},
\end{equation}
where $e_{\rm GW} = e_{\rm GW}(a)$, and we define the integral
\begin{equation}
    I(e_{\rm GW}) \equiv \int_0^1 x^{5/3}\left(1-e^2_{\rm GW}x^2\right)^{1/3}\left(1-x\right)^{-2/3}{\rm d}x.
\end{equation}
If most inspirals happen at $a\simeq a_{\rm GW}$, then
\begin{equation}
    \frac{\Sigma^{\rm ins}_{\rm d}}{\Sigma^{\rm ins}_{\rm r}}  \simeq 0.86 .
\end{equation}
This result is reasonably close to the ratio observed in our simulations, which is 2/3: indeed, both cross sections are of the same order of magnitude. From equation (\ref{eq sigma ratio}) and equation (\ref{eq lgw}) it can seen that the ratio of cross sections scales with $\Sigma^{\rm ins}_{\rm d} / \Sigma^{\rm ins}_{\rm r} \propto a^{10/7}$, where we neglect the weak dependence that $I(e_{\rm GW})$ has on $e_{\rm GW}$. This suggests that the relative importance of direct encounters increases for softer binaries.

\section{The effect of 1pN precession on direct inspirals}
\label{APPENDIX 1pn}

Here we estimate whether ignoring $1^{\rm st}$ order post-Newtonian terms is a good approximation for the specific case of direct inspirals. For that, we compare the timescale at which the BBH precesses, $\tau_{\omega}$, to the timescale at which the dimensionless angular momentum of the binary $l$ changes due to the field BH, $\tau_l$. If $\tau_l<\tau_{\omega}$, then it is safe to ignore relativistic precession. We can express both timescales as $\tau_{\omega}\sim\pi/|\dot{\omega}|$ and $\tau_l\sim l/|\dot{l}|$,
with 
\begin{equation}
    |\dot{\omega}| \simeq \frac{|\Delta\omega|}{T}=\frac{24\pi^3a^2}{T^3c^2l^2},
\end{equation}
where $T$ is the orbital period of the BBH, $\dot{\omega}$ the rate at which the perihelion shifts, and $\Delta\omega$ the perihelion shift during one orbital period. From this it follows that
\begin{equation}
    \tau_{\omega} \simeq \frac{T^3c^2l^2}{24\pi^2a^2}.
\end{equation}
We proceed now with $t_{l}$. We can express $|\dot{l}|$ as
\begin{equation}
    |\dot{l}| = \frac{e\dot{e}}{l} = \frac{e|\delta e|}{l\delta t},
\end{equation}
with $\delta e$ the change of eccentricity over a time $\delta t$. After \cite{heggie_rasio_1996}, $\delta e$ after a direct encounter is
\begin{equation}
    |\delta e| = \frac{15\pi}{16\sqrt{3}}el\left(\frac{a}{r}\right)^{3/2}{\rm sin}(2\Omega)\,{\rm sin}^2(i),
\end{equation}
where $\Omega$ is the longitude of the ascending node, and we assume equal masses. For the specific case of direct inspirals, the conditions are such that $\delta e$ is maximised, implying ${\rm sin}(2\Omega)\sim{\rm sin}(i)\sim1$. Using $l=\sqrt{1-e^2}$ we are left with
\begin{equation}
    |\dot{l}| \simeq \frac{17}{10}\left(\frac{a}{r}\right)^{-3/2}\frac{1-l^2}{\delta t}.
\end{equation}
We can now perform the ratio $\tau_l/\tau_{\omega}$, given by
\begin{equation}
    \frac{\tau_l}{\tau_{\omega}} \simeq \frac{120}{17}\frac{Gm}{c^2l(1-l^2)}\left(\frac{r}{a}\right)^3\frac{1}{a},
\end{equation}
where we use the fact that $\delta t$ scales with $r$ as $\delta t \sim T(r/a)^{3/2}$, and we use Kepler's third law to write $T$ in units $a$. Note that precession becomes increasingly important as $a$ shrinks. We place ourselves in the worst case scenario by adopting its minimum value $a=a_{\rm GW}$, given in Section \ref{sma of the binary}. For that, we use a cluster with $M_{\rm cl}=2\times10^5\,{\rm M}_{\odot}$, and $r_{\rm h}=3\,{\rm pc}$. In the particular case of direct inspirals, we assume $l$ to be close to the threshold $l_{\rm GW}$, given in Section \ref{inspirals}. Therefore
\begin{equation}
    \frac{\tau_l}{\tau_{\omega}} \simeq 1.6\times10^{-5} \left(\frac{r}{a}\right)^3.
\end{equation}
By imposing $\tau_l<\tau_{\omega}$ we obtain
\begin{equation}
    r < 18a.
\end{equation}
Consequently, even for the smallest possible SMA, precession is unimportant for all flybys with distance $r<18a$. In our scattering experiments, we found all direct inspirals to happen for $r<12a$ (see Fig.~\ref{ins_ecc}), therefore it is safe to ignore the effect of precession.
\end{appendix}

\end{document}